\documentclass[twocolumn]{aastex631}

\shorttitle{Coherent emission from QED cascades}
\shortauthors{Cruz et al.}
\graphicspath{{./}{figures/}}

\usepackage{amsmath}

\begin{document}

\title{Coherent emission from QED cascades in pulsar polar caps}

\correspondingauthor{F\'{a}bio Cruz}
\email{fabio.cruz@tecnico.ulisboa.pt}

\author[0000-0003-0761-6628]{F\'{a}bio Cruz}
\affiliation{GoLP/Instituto de Plasmas e Fus\~{a}o Nuclear, Instituto Superior T\'{e}cnico, Universidade de Lisboa, 1049-001 Lisboa, Portugal}

\author[0000-0002-0045-389X]{Thomas Grismayer}
\affiliation{GoLP/Instituto de Plasmas e Fus\~{a}o Nuclear, Instituto Superior T\'{e}cnico, Universidade de Lisboa, 1049-001 Lisboa, Portugal}

\author[0000-0002-4738-1168]{Alexander Y. Chen}
\affiliation{JILA, University of Colorado Boulder, 440 UCB, Boulder, CO 80309, USA}

\author[0000-0001-9179-9054]{Anatoly Spitkovsky}
\affiliation{Department of Astrophysical Sciences, Princeton University, Princeton, NJ 08544, USA}

\author[0000-0003-2906-924X]{Luis O. Silva}
\affiliation{GoLP/Instituto de Plasmas e Fus\~{a}o Nuclear, Instituto Superior T\'{e}cnico, Universidade de Lisboa, 1049-001 Lisboa, Portugal}



\begin{abstract}
Pulsar magnetospheres are thought to be filled with electron-positron plasma generated in pair cascades. The driving mechanism of these cascades is the emission of gamma-ray photons and their conversion into pairs via Quantum Electrodynamics (QED) processes. In this work, we present 2D particle-in-cell simulations of pair cascades in pulsar polar caps with realistic magnetic field geometry that include the relevant QED processes from first principles. Our results show that, due to variation of magnetic field curvature across the polar cap, pair production bursts self-consistently develop an inclination with respect to the local magnetic field that favors the generation of coherent electromagnetic modes with properties consistent with pulsar radio emission. We show that this emission is peaked along the magnetic axis and close to the polar cap edge and may thus offer an explanation for the core and conal components of pulsar radio emission.
\end{abstract}



\section{Introduction} \label{sec:intro}

Cascades of electron-positron pairs are a key source of plasma in pulsar magnetospheres. They result from a positive feedback loop that develops in vacuum gaps, regions of unscreened electric field in the magnetosphere. Polar caps have been proposed to host rotation-induced vacuum gaps~\citep{sturrock_1971, ruderman_sutherland_1975}, where TeV energy electrons and positrons emit gamma-ray photons via curvature radiation and these are absorbed in the local $\sim 10^{12}$G magnetic field to produce new pairs. Cascades cease when the fresh pair plasma screens the vacuum gap electric field. The plasma is then advected into the magnetosphere, the gap reopens and a new cascade begins~\citep[e.g.,][]{timokhin_2010}. The time-dependent dynamics of polar cap vacuum gaps has recently been proposed as a primary ingredient to explain the nature of pulsar radio emission~\citep{beloborodov_2008, timokhin_2013, philippov_2020, melrose_2020}.

Kinetic plasma simulations are ideally suited to study the highly nonlinear interplay between time-dependent pair cascades and coherent plasma processes. \citet{timokhin_2010} presented the first 1D particle-in-cell (PIC) simulations including curvature radiation and pair production from first principles. These simulations were performed in a frame corotating with the neutron star, which embeds the magnetic field twist imposed at macroscopic scales~\citep{bai_2010} via a background current density $j_\text{m} = \alpha \rho_\text{GJ} c$, where $\alpha$ is a constant of order unity, $\rho_\text{GJ}$ is the Goldreich-Julian~\citep{goldreich_julian_1969} charge density and $c$ is the speed of light. They showed that cascades develop over regular short bursts followed by long quiet phases during which no pairs are produced. As the gap is locally screened, electric field oscillations are inductively driven due to collective kinetic-scale plasma motions~\citep{levinson_2005, cruz_2021}. Similar simulations have been used by \citet{timokhin_2013, timokhin_2015} to determine the spectra and multiplicity of the pair plasma created in cascades for a variety of initial conditions.

\citet[][hereafter PTS20]{philippov_2020} have recently resorted to an heuristic description of the emission and pair production processes to perform the first 2D Cartesian PIC simulations of pair cascades and identified a new process of coherent radiation emission. The key ingredient for this process is a finite angle between the pair production front and the background magnetic field, which cannot be captured in 1D. While PTS20 have been able to demonstrate this coherent emission mechanism using a simplified configuration, the shape of the pair production front and the spectrum and Poynting flux profile of the emitted radiation in realistic field geometries depend on the microscopic details of the emission and pair production cross sections. First-principles simulations of polar cap pair cascades are very challenging computationally, due to i) the rapid and localized creation of a large number of particles and ii) the large separation between gap and plasma kinetic scales. While the polar cap vacuum gap extends for $\sim 100$~m~\citep{ruderman_sutherland_1975}, the electron skin depth associated with the dense plasma produced in the cascade is $\sim 1$~cm. This scale separation is a consequence of the large multiplicity of the cascade process that can be estimated as the ratio between the energies of primary and secondary particles, respectively $\varepsilon_\pm / m_e c^2 \sim 10^7$ and $\varepsilon_\pm' / m_e c^2 \sim 10^2$, where $m_e$ is the electron mass. In multidimensional configurations, the difficulties of simulating such scale disparity are heightened, and first-principles numerical models of the pulsar polar cap have not yet been possible.

In this Letter, we adopt a first-principles rescaling of the Quantum Electrodynamics (QED) processes responsible for gamma-ray and pair production processes in 2D axisymmetric PIC simulations of pulsar polar caps with a realistic field geometry. Using these simulations, we determine the multidimensional properties of pair cascades and their observational signatures.

\section{QED-PIC Simulations}
\label{sec:method}

We perform 2D QED-PIC simulations with the code OSIRIS~\citep{fonseca_2002, fonseca_2008} in axisymmetric cylindrical coordinates $(R, z)$. The lower $z$ boundary ($z = 0$) is a rotating conducting disk of radius $R_0$. The reference scale $R_0$ should be interpreted as the polar cap radius, and the $z$ axis as the magnetic (and rotation) axis of the neutron star. The rotation of the conducting disk is imposed by forcing the radial electric field at this boundary to be $E_R (R, z=0) = E_0 \times (R / R_0) \times g(R / R_0)$, where $E_0$ is a constant proportional to the angular velocity of the disk $\Omega$ and $g(x) = 0.5 \times (1 - \tanh( (x - 1) / 0.2 ))$ is a smooth cutoff function at $x \simeq 1$. This boundary condition induces a unipolar electric field near the conductor, forcing plasma in this region into corotation with the disk. The upper $z$ boundary is open for fields, such that any incident wave escapes freely from this boundary. Both $z$ boundaries are open for particles.

The simulation domain is also permeated by the externally imposed dipolar magnetic field $\mathbf{B_d}$ with magnitude $B_0$ at $R = z = 0$. The center of the magnetic dipole is at $(R, z) = (0, - R_*)$, with $R_* / R_0 = 10$. In analogy with a realistic polar cap, all magnetic field lines that cross the lower $z$ boundary at a radius $R < R_0$ are open to infinity, and those that cross this boundary at $R > R_0$ are closed on the other hemisphere of the neutron star. We assume that closed field lines are filled with dense plasma that we model as an ideal conductor, \textit{i.e.}, we set $\mathbf{E}_\parallel = (\mathbf{E} \cdot \mathbf{B_d}) \mathbf{B_d} / |\mathbf{B_d}|^2$ in this region to zero. This simulation setup forces a current to be driven along open field lines while providing a (virtual) return current along the edge of the conducting disk, mimicking the local conditions in pulsar polar caps.

The QED processes governing pair cascades in our simulations are photon emission via nonlinear Compton scattering~\citep{erber_1966}, the QED equivalent of curvature radiation for classical emission from ultra-relativistic particles~\citep{kelner_2015, delgaudio_2020}, and multiphoton Breit-Wheeler pair production~\citep{ritus_1985}. In our simulations, these processes are included using Monte-Carlo methods~\citep{grismayer_2016, grismayer_2017}. In each time step, we first compute the quantum parameter $\chi_{\pm, \gamma}$ of each particle (subscripts $\pm$, $\gamma$ correspond to electrons, positrons and photons, respectively), defined as $\chi_{\pm, \gamma} = \sqrt{(p_\mu F^{\mu \nu})^2} / (B_Q m_e c^2)$, where $p^\mu$ is the four-momentum of the particle, $F^{\mu \nu}$ is the electromagnetic tensor and $B_Q \simeq 4.4 \times 10^{13}$~G is the Schwinger field. We then rescale $\chi_\pm \to \chi_\pm' \equiv \zeta_\pm \chi_\pm$ and $\chi_\gamma \to \chi_\gamma' \equiv \zeta_\gamma \chi_\gamma$, with $\zeta_{\pm, \gamma} \gg 1$, and use $\chi_{\pm, \gamma}'$ to evaluate the probability of creating new particles. The rescaling is done to reduce the scale separation described above. A key aspect of our approach is that it retains the fundamental properties of the QED processes: i) the spectrum of photons emitted via nonlinear Compton scattering is preserved (in particular its dependence on the energy and curvature of the trajectory of the emitting particle), and ii) the probability of pair production critically depends on the angle between the photon propagation direction and the local electromagnetic field.

\begin{figure*}[ht]
\centering
\includegraphics[width=6.35in]{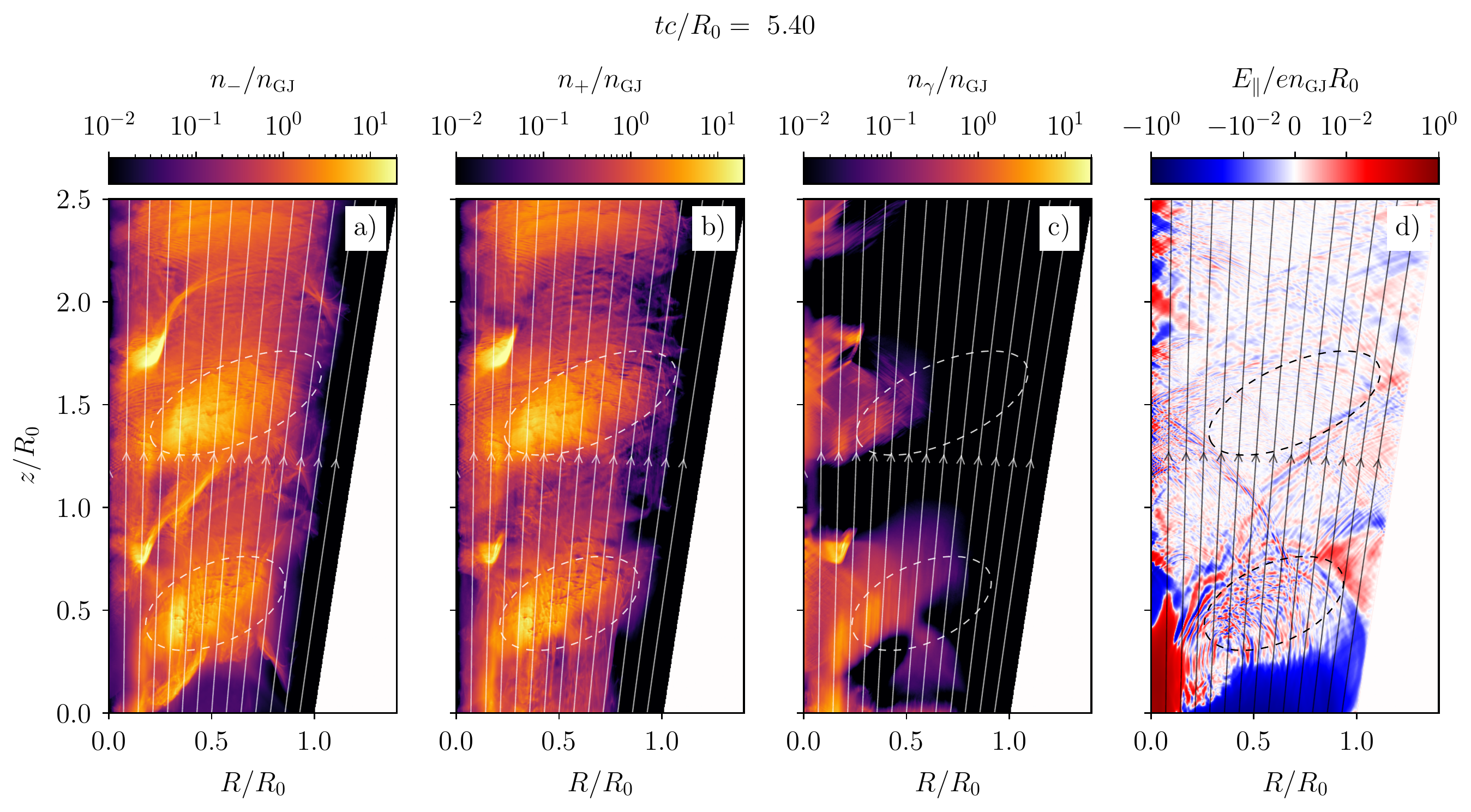}
\caption{\label{fig:den_ez_snapshot} Electron/positron/photon densities (a-c, respectively) and electric field component parallel to the background magnetic field (d) at a time $t c / R_0 = 5.4$, where two bursts are visible. All panels show the closed field line region displayed in white and the magnetic field lines in solid (white in a-c, black in d) lines. A visual aid identifying the pair production bursts (\textit{i.e.}, outlining large density regions) is shown in dashed lines in all panels.}
\end{figure*}

The surface electric field $E_0 = \Omega B_0 R_* / c$ and the rescaling parameters $\zeta_{\pm, \gamma}$ are adjusted such that a lepton is accelerated to the radiation reaction limited Lorentz factor~\citep{daugherty_1982}
\begin{equation}
\hat{\gamma}_\pm = \left( \frac{3}{2\alpha_\text{fs}} \frac{E_0}{B_Q} \right)^{1/4} \left( \frac{\lambdabar_\text{C}}{\rho} \right)^{-1/2} \zeta_\pm^{-1/2}
\label{eq:gamma_hat}
\end{equation}
over a distance 
\begin{equation}
\ell_a \simeq \hat{\gamma}_\pm \left(\frac{B_Q}{E_0}\right) \lambdabar_\text{C} \ ,
\label{eq:ell_a}
\end{equation}
that we take to be $\sim 0.1 R_0$. In equations~\eqref{eq:gamma_hat} and \eqref{eq:ell_a}, $\alpha_\text{fs}$ is the fine-structure constant, $\lambdabar_\text{C}$ is the reduced Compton wavelength and $\rho \simeq R_*^2 / R_0$ is the radius of curvature of the last open field line. We fix $\zeta_{\pm, \gamma} = 10^3$, $e B_0 R_0 / m_e c^2 = 10^8$ and $B_0 / B_Q = 0.1$, and vary only $E_0$ such that $\ell_a / R_0 \simeq \{0.2, 0.35, 0.5, 0.6\}$. These values were chosen for simulations to be computationally feasible yet comparable to real pulsars, in which $\ell_a / R_0$ varies from $\sim 10^{-4}$ to $\sim 1$ with increasing rotation period. The mean free path of photons emitted at the critical energy $\varepsilon_c = (3/2) (\lambdabar_\text{C}/\rho) \hat{\gamma}_\pm^3 m_e c^2 \zeta_\pm$, defined as~\citep[e.g.,][]{timokhin_2015} 
\begin{equation}
\ell_\gamma \simeq \frac{8}{3} \frac{B_Q}{B} \frac{m_e c^2}{\varepsilon_c} \rho \zeta_\gamma^{-1} \ ,
\label{eq:ell_gamma}
\end{equation}
is always $\sim 10^{-2}~R_0$ for our choice of parameters. Thus, the ratio $\ell_\gamma / \ell_a \sim 0.1$ is also within the range $\sim 10^{-2} - 1$ expected in real systems. The electron skin depth associated with a density $n_\text{GJ} = |\rho_\text{GJ}| / e \equiv \Omega B_0 / 2 \pi c e$ is $\sim 5 \times 10^{-3} - 10^{-2}~R_0$. The simulation domain has a size $L_R \times L_z = (1.5R_0) \times (2.5R_0)$ discretized in $N_R \times N_z = 6000 \times 10000$ cells, and the time step is $\Delta t = 10^{-4}R_0 / c$. The grid size was chosen to resolve the electron skin depth associated with the maximum density generated during pair cascades.

The rotating conducting disk is gradually spun up to its maximum angular velocity in the first 200 time steps of the simulations and kept constant thereafter. There is initially no plasma in the simulation domain, so this spin induces a vacuum corotation electric field in the open field line region: $E_R$ increases and $E_z$ decreases with $R$, whereas both components decay with $z$ within a distance $\sim R_0$. We let the corotation field develop for a light-crossing time, $L_z / c = 2.5 ~ R_0 / c$, and then start injecting plasma in cells just above the $z = 0$ boundary for all $R < R_0$. At every time step, we inject one electron-positron pairs per cell carrying a density $n_\text{inj} = \kappa (E_\parallel / e R_0)$, where $\kappa = 0.1$ and $E_\parallel = |\mathbf{E}_\parallel|$. Pairs are injected at rest. We choose a value of $\kappa \ll 1$ to ensure this injection provides only a seed plasma for the cascades and does not dominate the plasma outflow.

\section{Results}
\label{sec:results}

As pairs are injected and experience the vacuum electric field, only electrons are able to escape the surface of the conducting disk, whereas positrons are immediately reabsorbed at the boundary. In their acceleration along the magnetic field lines, electrons emit curvature photons, which then decay into pairs, triggering the pair cascade. The accelerating electric field $E_\parallel$ is then locally screened and the cascade stops. When the plasma flows away from the conducting disk, the vacuum gap reopens and the process restarts.

In Figure~\ref{fig:den_ez_snapshot}, we show the electron, positron and photon densities and $E_\parallel$ for a simulation with $\ell_a / R_0 \simeq 0.2$ when two pair production bursts are visible in the simulation domain. Due to horizontal gradients of the magnetic field curvature, photon emission and pair production do not occur uniformly across the open field lines. Instead, the cascade is triggered at $R \simeq R_0 / 2$ and $z \simeq \ell_a$. For $R \gtrsim R_0 / 2$, $E_\parallel$ decreases, limiting acceleration and consequently photon emission, whereas for $R / R_0 \ll 1$ the radius of curvature of magnetic field lines rapidly diverges and pair production is suppressed~\citep{arons_1979}. For $R \lesssim R_0 /2$, the pair cascade develops at $z < \ell_a$, \textit{i.e.}, the pair production front is inclined towards the magnetic axis.

As pair production bursts develop, inductive plasma waves are self-consistently excited~\citep{levinson_2005, cruz_2021}. Due to the two-dimensional structure of the pair production front, these waves are not only emitted in the positive $z$ direction, but also in all other directions --- see Figure~\ref{fig:den_ez_snapshot}d. The inclination of the wavevectors relative to $\mathbf{B}$ is not the result of cross-field particle motion, but rather of field-aligned currents coherently developed on adjacent field lines.

This inclination is a key ingredient in the self-consistent excitation of electromagnetic plasma modes. The wave generation process can be understood as follows~\citep[PTS20, ][]{melrose_2020}: first, the non-uniformity of $\mathbf{E}_\parallel$ across dipolar field lines gives rise to an oscillating azimuthal component of the magnetic field $\tilde{B}_\phi$ via $(\partial \mathbf{\tilde{B}} / \partial t) \sim - c \nabla \times \mathbf{E}_\parallel$, where $\mathbf{\tilde{B}} = \tilde{B}_\phi \mathbf{e}_{\boldsymbol{\phi}}$; then, an oscillatory electric field component $\mathbf{\tilde{E}}$ is excited via $(\partial \mathbf{\tilde{E}} / \partial t) \sim c \nabla \times \mathbf{\tilde{B}}$. Gradients in $\mathbf{\tilde{B}}$ occur predominantly along the normal to the pair production front, which is also the direction of the wavevector $\mathbf{k}$ of these waves. In general, both $\mathbf{k}$ and $\mathbf{\tilde{E}}$ have components parallel and perpendicular to $\mathbf{B}_\textrm{d}$. The requirement that the angle between the normal to the pair production front and $\mathbf{B}_\textrm{d}$ is finite is essential for this process to operate. Hence, modes with $\mathbf{k} \parallel \mathbf{B}_\textrm{d}$ are never produced without accompanying modes with $\mathbf{k} \perp \mathbf{B}_\textrm{d}$. All our simulations show that these modes are excited at the local plasma frequency $\omega_0$ (\textit{i.e.}, at the frequency of inductive $\mathbf{E}_\parallel$ oscillations) and are linearly polarized. As pairs are produced, the local density increases and the wave frequency extends to $\sim 10-100\omega_0$. The properties of these modes are consistent with the superluminal O-modes\footnote{This mode should not be confused with the O-mode presented in plasma physics textbooks~\citep[e.g.,][]{nicholson_1983, stix_1992}. The mode presented here propagates in a range of directions from purely perpendicular to purely parallel to $\mathbf{B}_\textrm{d}$, while textbook O-modes exist only for $\mathbf{k} \perp \mathbf{B}_\textrm{d}$.} identified in previous works~\citep[][PTS20]{arons_1986}.

\begin{figure}
\centering
\includegraphics[width=3.375in]{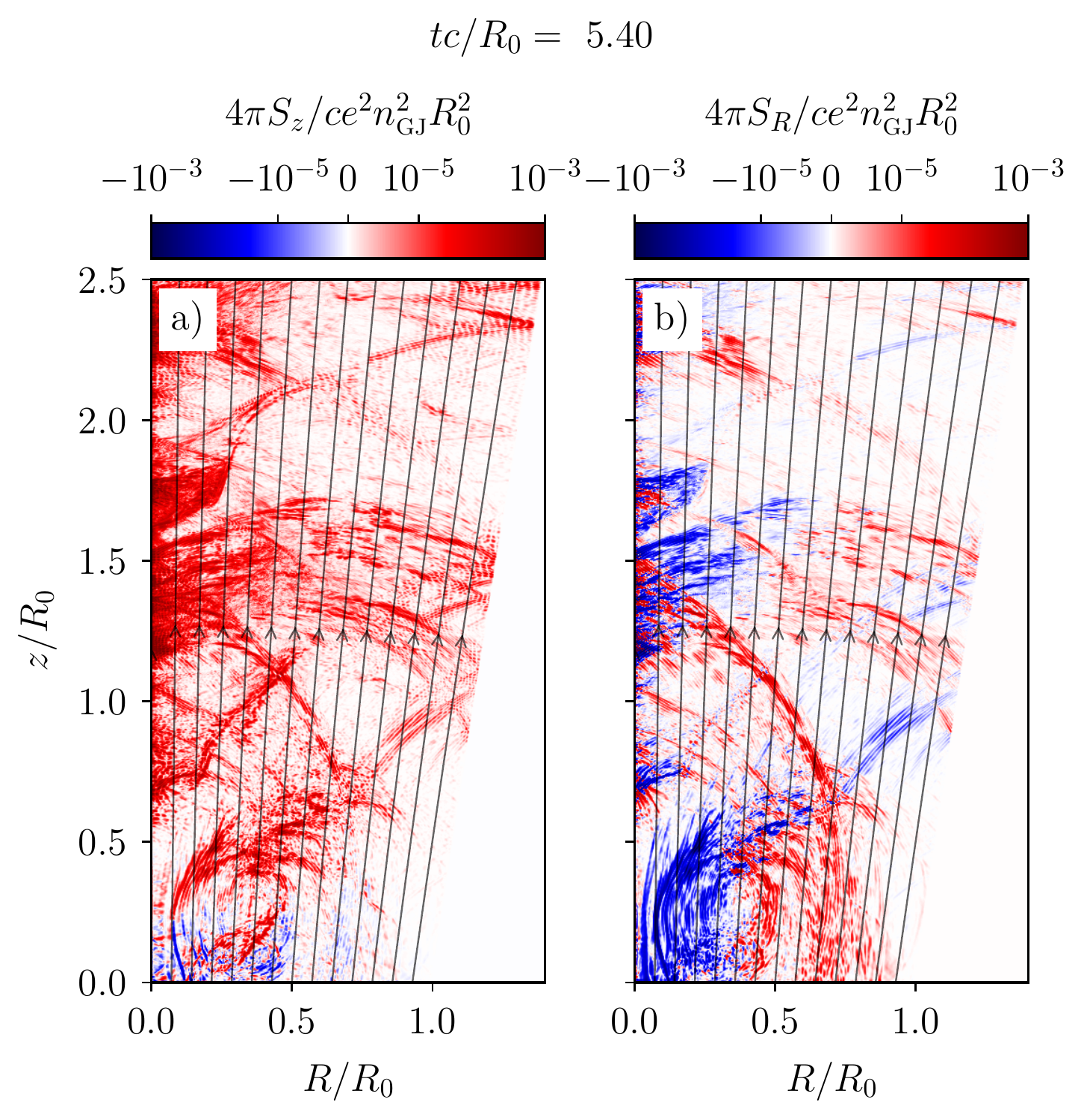}
\caption{\label{fig:sz_sr_snapshot} Poynting vector components $S_z$ and $S_R$ associated with the fluctuating electromagnetic field components (calculated by subtracting a local average to the total fields). This snapshot was taken at the same time as Figure~\ref{fig:den_ez_snapshot}. Red (blue) tones identify regions where there is a positive (negative) electromagnetic energy density flux in the $z$ and $R$ directions in panels a and b, respectively. The magnetic field lines are shown in solid black lines.}
\end{figure}

A snapshot of the components of the Poynting vector $\mathbf{S} = (c / 4 \pi) \mathbf{\tilde{E}} \times \mathbf{\tilde{B}}$ is shown in Figure~\ref{fig:sz_sr_snapshot}. The wave components $\mathbf{\tilde{E}}$ and $\mathbf{\tilde{B}}$ were calculated by subtracting local time-averaged $\mathbf{E}$ and $\mathbf{B}$, respectively. Figure~\ref{fig:sz_sr_snapshot}a shows rings of electromagnetic flux emitted around the most recently created plasma burst, centered at $z / R_0 \simeq 0.4$ and $R / R_0 \simeq 0.5$. Waves emitted upward carry a positive $S_z$, whereas waves emitted downward have initially a negative $S_z$ that is then reversed after the waves are reflected at the $z = 0$ boundary. Electromagnetic waves are emitted in all directions. However, due to the inclination of the bursts relative to the magnetic field lines, a larger Poynting flux is generated on the flanks of the bursts. After some altitude, $S_z$ thus exhibits a double-peaked structure, being larger close to the magnetic axis and the conducting field line boundary and smaller in the center of the open field line bundle. 

\begin{figure*}[t]
\centering
\includegraphics[width=6.35in]{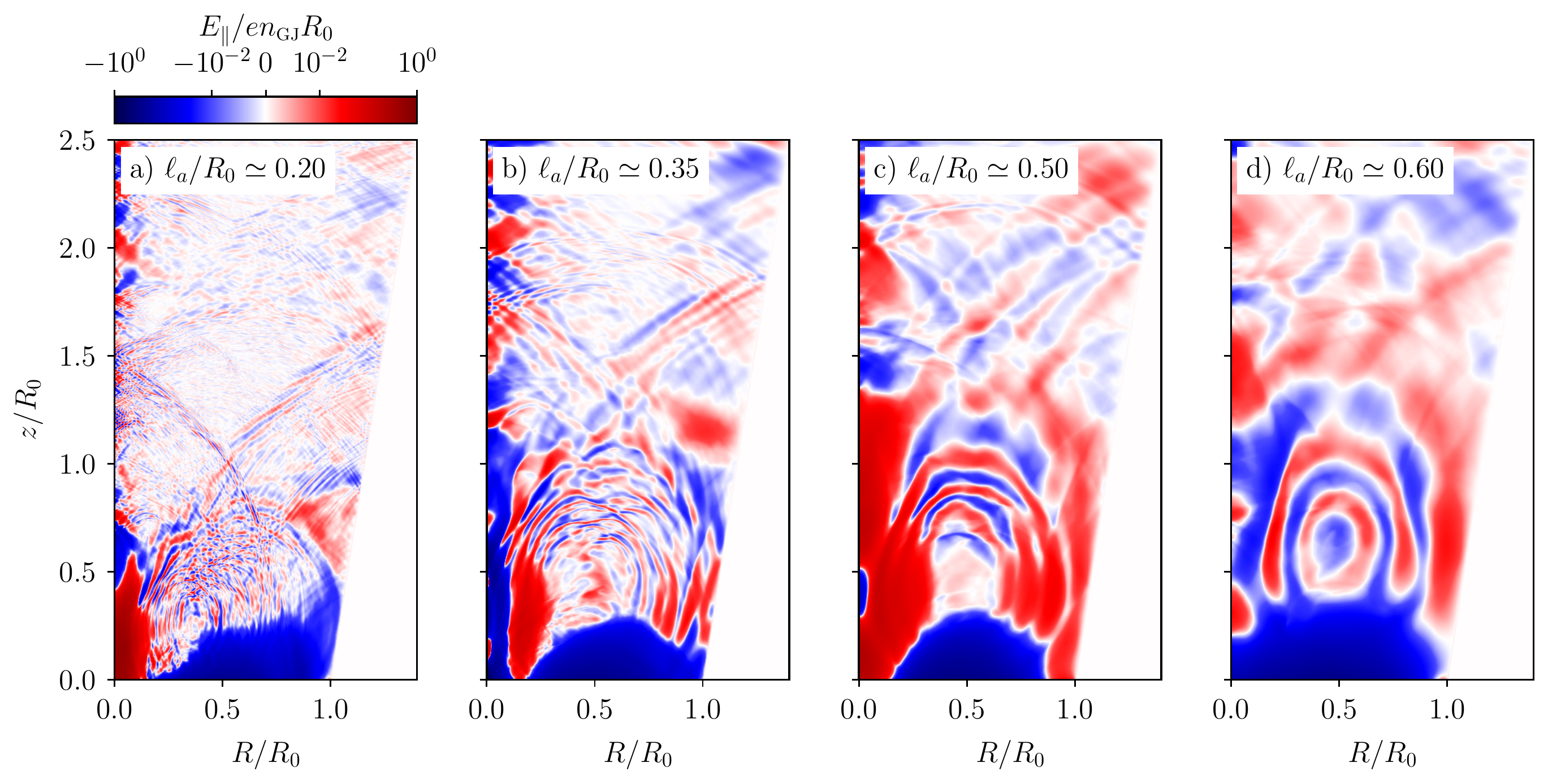}
\caption{\label{fig:epar_comp} Comparison between $E_\parallel$ oscillations induced by cascades with different $\ell_a / R_0$, where $\ell_a$ is the characteristic distance required for electrons to be accelerated to energies capable of emitting pair producing photons. Both the wavelength of the oscillations and the aspect ratio of the pair production burst (given by the ratio between its height and width) increase with $\ell_a / R_0$. The snapshots were taken at times $t = 5.4$, $5.25$, $5.75$ and $6.25 R_0 / c$ for $\ell_a/R_0 = 0.2$, $0.35$, $0.5$ and $0.6$, respectively. A movie showing the temporal evolution of these quantities is \href{https://youtu.be/wHSw5kp7Ik4}{available online.}}
\end{figure*}

Figure~\ref{fig:sz_sr_snapshot}b shows that the part of the electromagnetic flux rings convert into bands extended in the $z$ direction (see e.g. $R / R_0 \simeq 0.5$ and $z / R_0 \simeq 1$) over time. These bands result from the continuous reflection of the waves between the magnetic axis and the conducting region of closed field lines, and they drift across the open field line region over a time scale $\sim R_0 / c$.

\begin{figure}[t]
\centering
\includegraphics[width=3.35in]{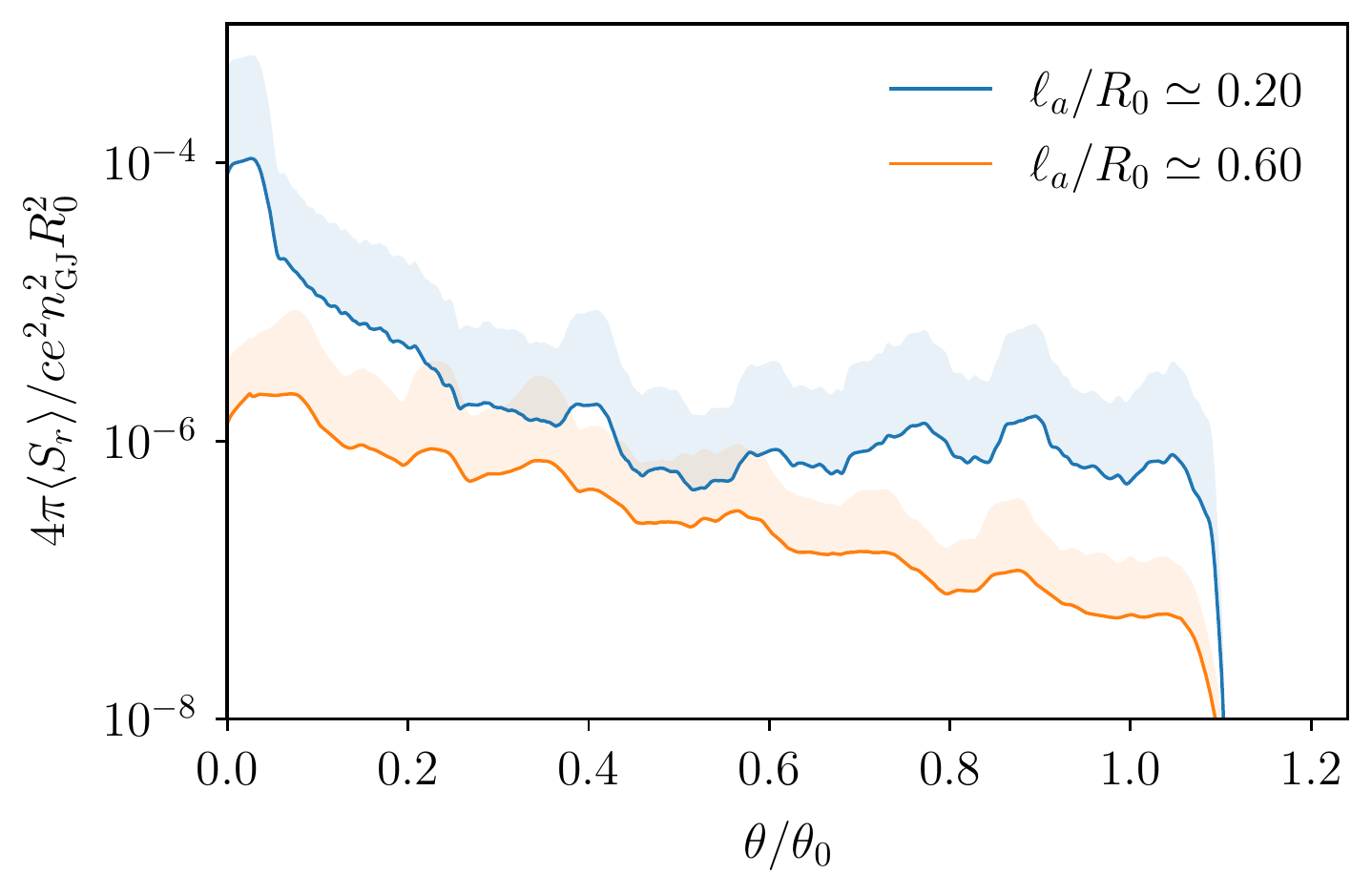}
\caption{\label{fig:sr_comp} Average Poynting flux $\langle S_r \rangle$ at $z / R_0 = 2$ for simulations with $\ell_a / R_0 = 0.2$ (blue) and $\ell_a / R_0 = 0.6$ (orange). The shaded regions above each curve extend for a standard deviation above the time average.}
\end{figure}

The electrodynamics of the two-dimensional pair cascades identified above for $\ell_a / R_0 \simeq 0.2$ holds also for different values of $\ell_a / R_0$. However, there are important effects to note: first, the wavelength of the $E_\parallel$ oscillations (and consequently of $\mathbf{\tilde{E}}$ and $\mathbf{\tilde{B}}$) decreases with decreasing $\ell_a$, due to the larger multiplicity generated in the cascades in this regime; second, the pair production burst extends to smaller $\theta$ for smaller $\ell_a / R_0$; third, the shape of the pair production burst is flatter in the radial direction for smaller $\ell_a / R_0$, and more round for larger $\ell_a / R_0$ --- see Figure~\ref{fig:epar_comp}. The latter effect is responsible for a more efficient generation of Poynting flux, since it allows for a larger portion of the pair production bursts to excite waves with $\mathbf{k}$ almost purely aligned with $z$. This is visible in Figure~\ref{fig:sr_comp}, where we show $\langle S_r \rangle \equiv \langle S_z \cos \theta + S_R \sin \theta \rangle$, averaged for all times after the initial vacuum transient at $z / R_0 = 2$ for simulations performed with different values of $\ell_a / R_0$. The angle $\theta$ is normalized to $\theta_0 \equiv \arcsin (R_0 / R_*)$, \textit{i.e.}, the analogue of the polar cap angle in pulsar magnetospheres. For $\ell_a / R_0 \simeq 0.2$, we observe the double-peaked structure in this profile described before (see peaks at $\theta \simeq 0$ and $\theta \simeq \theta_0$). However, for large $\ell_a / R_0$, the edge component is absent. We also observe that the peak value of $\langle S_r \rangle$ on the magnetic axis decreases with $\ell_a / R_0$, but is always $\sim 10^{-6} - 10^{-4}~S_0$, where $S_0 = c e^2 n_\text{GJ}^2 R_0^2 / 16 \pi$ is the total Poynting flux launched by the rotating conductor, a result consistent with the fraction of pulsar spin down power observed in the radio~\citep[e.g,][]{lorimer_kramer_2004}.

\section{Discussion}
\label{sec:discussion}

In this Letter, we have presented the first 2D simulations of pulsar polar cap pair cascades including the QED effects from first principles. Our results show that the gap dynamics has two significant two-dimensional features that could not be captured in one-dimensional simulations: a) pair production is inhibited close to the magnetic axis, due to the null curvature of the magnetic field in this region, and b) gradients of the magnetic field curvature across the polar cap induce an inclination between the normal to the pair production front and the background magnetic field.

We have observed that $E_\parallel$ oscillations inductively driven in the inclined pair production bursts can act as a source for coherent electromagnetic waves. Although the initial oscillations in $E_\parallel$ is longitudinal, the induced waves are oblique and electromagnetic in nature, which makes them distinct from the L mode waves discussed in previous works~\citep{rafat_2019, melrose_2020}. Moreover, the electromagnetic modes observed in our simulations are naturally produced in the oblique pair production fronts, and do not require the production of intermediary modes such as the L modes suggested by~\citet{melrose_2020}. The induced electromagnetic waves propagate in all directions, but the Poynting flux flows predominantly outwards, away from the star. We have identified the coherent electromagnetic modes to be linearly polarized, superluminal O-modes, and verified that the resulting Poynting flux from this emission has two peaks: one on the magnetic axis and another on the edge of the open field line bundle. We interpret this as a consequence of the relative orientation between the normal to the pair production bursts and the background magnetic field. Thus, we expect this mechanism to operate also in magnetic field topologies more complex than a pure dipole (e.g. in multipolar magnetic fields), provided that the field curvature changes over a scale larger than the gap height $\sim \ell_a$. In such topologies, the configuration of the open field lines, where cascades develop, may lead to complex-shaped pair production bursts and thus different emission power profiles.

We have also shown that a fraction of the emitted waves is continuously reflected at last open field line boundary, giving rise to a drifting component in the Poynting flux on a time scale $\sim 1$~$\mu$s. We note that this should not be confused with the observed radio drifting sub-pulses, that occur on time scales $\sim 0.1-1$~s~\citep[e.g.,][]{weltevrede_2007}. The drifts identified in our simulations should be detectable at high time resolution and could be used to diagnose the height of the emission.

The simulations presented in this work adopt a very dilute, space charge-limited flow from a rotating conductor, and focus on the role of pair cascades in providing the current to screen the vacuum gap. In reality, the neutron star is expected to provide a larger current density $\sim j_\text{GJ} = \rho_\text{GJ} c$. In that case, the vacuum gaps are induced by the inability of the star to match the current density $j > j_\text{GJ}$, required by the global magnetosphere when general relativistic frame-dragging effects are taken into account~\citep{gralla_2016}. It is not expected that pair cascades operate differently when these effects are included; however, given the same $B_0$ and $\Omega$, the ratio $\ell_a / R_0$ may differ slightly from what is presented in this work.


\vspace{0.4cm}
FC, TG and LOS are supported by the European Research Council (ERC-2015-AdG Grant 695088). FC is also supported by FCT (Portugal) (grant PD/BD/114307/2016) in the framework of the Advanced Program in Plasma Science and Engineering (APPLAuSE, FCT grant PD/00505/2012). AC is supported by NSF grants AST-1806084 and AST-1903335. AS is supported by NASA grant 80NSSC18K1099 and NSF grant PHY-1804048. We acknowledge PRACE for granting access to MareNostrum (Barcelona Supercomputing Center, Spain) and to TGCC Joliot Curie (CEA, France), where the simulations presented in this work were performed.


\bibliography{sample631}{}
\bibliographystyle{aasjournal}



\end{document}